# Universal relation between magnetic resonance and superconducting gap in unconventional superconductors

G. Yu[1], Y. Li[1], E. M. Motoyama[1] and M. Greven[2*]

[1]Department of Physics, Stanford University, Stanford, California 94305,
[2]Department of Applied Physics, Stanford University, Stanford, California 94305.
*e-mail: greven@stanford.edu.

**Unconventional superconductors such as the high-transition temperature ($T_c$) cuprates, heavy-fermion systems and iron arsenide-based compounds exhibit antiferromagnetic fluctuations that are dominated by a resonance, a collective spin-one excitation mode in the superconducting state[1-5]. Here we demonstrate the existence of a universal linear relation, $E_r \propto 2\Delta$, between the magnetic resonance energy ($E_r$) and the superconducting pairing gap ($\Delta$), spanning two orders of magnitude in energy. This relation is valid for materials that range from being close to the Mott-insulating limit to being on the border of itinerant magnetism. Since the common excitonic picture of the resonance has not led to such universality, our observation suggests a much deeper connection between antiferromagnetic fluctuations and unconventional superconductivity.**

Superconductivity involves the formation of electron pairs (Cooper pairs) and their condensation into a macroscopic quantum state. In conventional superconductors, such as $Nb_3Ge$ and elemental Hg, weakly-interacting electrons pair via the electron-phonon interaction, forming spin singlets described by an isotropic (*s*-wave) gap function. In contrast, unconventional superconductivity appears in correlated-electron materials in which electronic interactions are significant and the pairing mechanism may not be phononic. In the high-$T_c$ cuprates, the superconductivity arises upon doping charge carriers into the copper-oxygen layers of antiferromagnetic (AF) Mott insulators[6]. Other examples of unconventional superconductors are the heavy-fermion compounds, which are metals with coupled conduction and localized f-shell electrons[7], and the recently discovered iron arsenide superconductors[8]. These unconventional superconductors exhibit anisotropic gaps and/or multiple bands at the Fermi level. There exist stark differences in the phase diagrams of these three classes of materials, yet they all exhibit significant AF correlations that may contribute to or be the cause of the superconductivity. Further clarification of the relationship between magnetism and superconductivity appears pivotal to resolving the underlying mechanisms.

The most prominent feature in the magnetic excitation spectrum in the superconducting (SC) state of the cuprates is the magnetic resonance, a collective mode at the twodimensional AF wave vector $(\pi/a, \pi/a)$[1-2]. Early inelastic neutron scattering experiments for $YBa_2Cu_3O_{6+\delta}$ (YBCO) indicate that the resonance energy has an approximately linear dependence on $T_c$: $E_r = 5\text{-}6\, k_B T_c$. However, the crystal structure of YBCO contains two nearby Cu-O layers per unit cell, and later experiments revealed at higher energy an even-parity resonance that differs from the previously observed odd-parity mode in its symmetry with respect to the exchange of the two layers[9-11]. The resonance is also observed in neutron scattering measurements of other hole-doped cuprates: double-layer $Bi_2Sr_2CaCu_2O_{8+\delta}$ (Bi2212)[12-13], and the single-layer systems $Tl_2Ba_2CuO_{6+\delta}$ (Tl2201)[14] and $HgBa_2CuO_{4+\delta}$ (Hg1201)[15-16], which exhibit only one resonance mode. Investigation of $Pr_{1-x}LaCe_xCuO_{4+\delta}$ (PLCCO)[17] and $Nd_{2-x}Ce_xCuO_{4+\delta}$ (NCCO)[18] suggests that the resonance may exist in the electron-doped cuprates as well. Resonance-like magnetic peaks at the AF zone center were also found in the SC state of the heavy-fermion superconductors $UPd_2Al_3$[3] and $CeCoIn_5$[4], as well as in the iron arsenide $(Ba,K)Fe_2As_2$[5], indicating that the resonance is a rather universal feature in the magnetic spectra of unconventional superconductors.

For the cuprates, the mode energy $E_r$ of the resonance is commonly compared with $T_c$. Figure 1 shows that the linear correlation $E_r = 5\text{-}6\ k_BT_c$ is approximately satisfied for the odd-parity mode of the double-layer compounds YBCO and Bi2212, for the resonance in Tl2201, and even for the characteristic energy of the local susceptibility in $La_{2-x}Sr_xCuO_4$ (LSCO)[19-20], for which no resonance has been observed. However, recent experiments reveal a violation of the proportionality between $E_r$ and $T_c$ for single-layer Hg1201[15-16] and for electron-doped NCCO[18]. Furthermore, below optimal doping (hole concentration $p < 0.16$), the odd and even resonance energies of the double-layer compounds differ significantly. The average of the two is considerably larger than $6\ k_BT_c$ for the most underdoped YBCO samples in which both modes have been studied[11,21] (Fig. 1).

The characteristic energy scale of superconductivity is the SC gap $\Delta$ in the low-energy single-particle response. As seen from Fig. 2, our analysis shows that $E_r$ is universally related to $\Delta$ rather than to $T_c$. We note that for the hole-doped cuprates, the determination of the SC gap is complicated by the existence of multiple energy scales[22], the deviation of the gap function from a pure $d$-wave form[23], and the spatial inhomogeneity of the gap seen in scanning tunneling microscopy (STM) experiments[24]. From photoemission measurements, the SC gap is associated with the gap in the low-energy charge excitations near the nodal region in momentum space, whereas a distinct 'pseudogap' is observed in the antinodal region. The latter is associated with many unusual charge and magnetic properties in the normal state[25]. In the underdoped regime, we estimate the SC gap maximum $\Delta$ from the nodal gap $\Delta_0$ determined in recent photoemission experiments. Above optimal doping, we estimate $\Delta$ from tunneling, thermal conductivity, and the spatially averaged gap determined in recent STM measurements. The details of our estimation of $\Delta$ in the hole-doped cuprates are found in the Supplementary Information. For $(Ba,K)Fe_2As_2$, two SC gaps are observed on different pieces of Fermi surface. We consider the larger of the two, which is associated with nesting bands that are thought to be responsible for the magnetic resonance. For all of the other systems, the SC gap is obtained directly from the literature.

We estimate $E_r/2\Delta = 0.64(2)$ from a fit to the combined result (Fig. 2), which spans two orders of magnitude in energy. A possible connection between the resonance energy and the SC gap has been previously suggested based on observations for the heavy-fermion compound $CeCoIn_5$[4]. Here we emphasize the universality of this connection across various classes of the unconventional superconductors and, in particular, we arrive at a consistent picture for the entire family of cuprates. Note that for the double-layer hole-doped cuprates, the resonance energy we consider is the average of even and odd parity modes.

The magnetic resonance appears to be associated with a SC gap function that undergoes a sign change. This is naturally the case in a Fermi-liquid picture for the *d*-wave cuprates, since the singularity of the magnetic susceptibility due to the coherence factor appears at momenta $\mathbf{Q} = (\pi/a, \pi/a)$ that satisfy $\Delta(\mathbf{q} + \mathbf{Q}) = -\Delta(\mathbf{q})$[2]. We note that although weak-coupling theory might be appropriate for the very overdoped part of the cuprate phase diagram, it is inadequate at lower doping[6]. The heavy-fermion superconductors considered here also appear to exhibit a *d*-wave order parameter[7]. On the other hand, the SC gap of the iron arsenide is believed to be nodeless. Nevertheless, the existence of the resonance may be accounted for by the possible anti-phase correlation between nesting hole- and electron-pockets responsible for the magnetic excitations[5]. This is further discussed in the Supplementary Information.

The significance of the linear scaling relation demonstrated in Fig. 2 lies in its simplicity and its universality among different types of unconventional superconductors, which has not been predicted theoretically. In some theoretical scenarios, the resonance is considered to be a π-mode, due to staggered *d*-wave particle-particle charge ±2 excitations, or a magnetic plasmon, as a result of mixing between the spin and charge channels[26]. The most widely considered theoretical scenario views the resonance as a spin exciton: a spin-1 particle-hole excitation with momentum $\mathbf{Q} = (\pi/a, \pi/a)$ that is bound at an energy below the pair-breaking energy (i.e., $E_r < 2\Delta$)[2]. The energy difference $2\Delta - E_r$ depends on the specific interactions (e.g., spin or charge fluctuations) stabilizing the bound particle-hole pair and is expected to differ among different classes of superconductors, and for different doping regimes of the hole-doped cuprates. Applied to the hole-doped cuprates, random phase approximation (RPA) approaches to the excitonic picture generally predict the ratio $E_r/2\Delta$ to be near 1 for the very overdoped regime, and to significantly decrease toward the underdoped side, rather than to take on a universal value[2]. Such approaches assume weak coupling and are thus expected to break down in the underdoped regime. A recent strong-coupling calculation for the Hubbard model near optimal doping predicts a significant increase of the resonance energy $E_r$ with doping, consistent with the behavior of the odd-parity mode in underdoped YBCO, but inconsistent with the doping independence of $E_r$ below optimal doping for single-layer Hg1201[15-16] and for the average mode energy of YBCO[27].

One scenario for the unconventional superconductivity is that it is mediated by magnetic fluctuations. In that case, the SC energy scale Δ would naturally be expected to follow from magnetic energy scales. Such a scenario implies that the electron-boson spectral function, which contains information about the 'paring glue' for superconductivity, should bear a clear signature of the magnetic spectra. For the cuprates, estimates of the electron-boson spectral function have been obtained by inverting optical spectra [28-29]. One approach[28] has revealed a well-defined peak in the SC state at a characteristic energy $E_{opt}$ that agrees remarkably well with $E_r$ (Fig. 2b), although for the double-layer compounds Bi2212 and YBCO, $E_{opt}$ appears to be dominated by the odd-parity resonance mode. For triple-layer Hg1223, for which no neutron data are yet available, $E_{opt} \approx 72$ meV[28] agrees well with the scaling established in Fig. 2.

It has been argued that the magnetic resonance itself has relatively small spectral weight[30] and that magnetic fluctuations throughout the whole Brillouin zone and over an extended energy range may be relevant to superconductivity[31]. Optical spectroscopy is an inherently momentum-integrated probe, and thus the extracted electron-boson spectral function is related to a momentum-integral of the magnetic susceptibility $\chi''(\mathbf{Q},\omega)$ rather than the value at the specific momentum $\mathbf{Q} = (\pi/a, \pi/a)$. For LSCO, it is the characteristic energy of the local (momentum-integrated) susceptibility $\chi''(\omega)$ that is shown in Figs. 1 and 2. For CeCoIn$_5$[4], the resonance is very prominent, so that the characteristic energy of $\chi''(\omega)$ is expected to be close to the value of the resonance energy. For NCCO, the characteristic energies of the peak and local susceptibilities are nearly indistinguishable as well[18]. Consequently, it is possible that the lowest characteristic energy of $\chi''(\omega)$ also scales with the SC gap.

The observation of a universal connection between the magnetic resonance and the SC gap for a wide range of materials seems to indicate that magnetic excitations play an important role in the Cooper-pair formation. To fully understand the connection between the magnetic resonance, the local magnetic susceptibility, and the electron-boson spectral function, more experimental information from inelastic neutron scattering is needed. Measurements of the doping and temperature dependence of the local magnetic susceptibility of single-layer Hg1201 and of the combined even and odd parity excitations in double-layer YBCO and Bi2212 are particularly desirable.

**Acknowledgements**
We thank A. V. Chubukov, K. K. Gomes, R.-H. He, S. A. Kivelson and A.-M. Tremblay for helpful comments. This work was supported by the US DOE under Contract No. DE-AC02-76SF00515 and by the NSF under Grant. No. DMR-0705086.

**References**

1. Rossat-Mignod, J. et al. Neutron scattering study of the $YBa_2Cu_3O_{6+x}$ system. *Physica C* **185,** 86 (1991).
2. Eschrig, M. The effect of collective spin-1 excitations on electronic spectra in high-$T_c$ superconductors. *Adv. in Phys.* **55,** 47 (2006).
3. Sato, N. K. et al. Strong coupling between local moments and superconducting 'heavy' electrons in $UPd_2Al_3$. *Nature* **410,** 340 (2001).
4. Stock, C. et al. Spin resonance in the $d$-Wave superconductor $CeCoIn_5$. *Phys. Rev. Lett.* **100,** 087001 (2008).
5. Christianson, A. D. et al. Unconventional superconductivity in $Ba_{0.6}K_{0.4}Fe_2As_2$ from inelastic neutron scattering. *Nature* **456,** 930 (2008).
6. Lee, P. A., Nagaosa, N. & Wen X. Doping a Mott insulator: Physics of high-temperature superconductivity. *Rev. Mod. Phys.* **78,** 17 (2006).
7. Goll, G. *Unconventional Superconductors: Experimental Investigation of the Order-Parameter Symmetry* (Springer, 2005).
8. Kamihara, Y. et al. Iron-based layered superconductor $La[O_{1-x}F_x]FeAs$ ($x$ = 0.05-0.12) with $T_c$ = 26 K. *J. Am. Chem. Soc.* **130,** 3296 (2008).
9. Pailhès, S. et al. Two resonant magnetic modes in an overdoped high $T_c$ superconductor. *Phys. Rev. Lett.* **91,** 237002 (2003).
10. Pailhès, S. et al. Resonant magnetic excitations at high energy in superconducting $YBa_2Cu_3O_{6.85}$. *Phys. Rev. Lett.* **93,** 167001 (2004).
11. Pailhès, S. et al. Doping dependence of bilayer resonant spin excitations in $(Y,Ca)Ba_2Cu_3O_{6+x}$. *Phys. Rev. Lett.* **96,** 257001 (2006).
12. Fong, H. F. et al. Neutron scattering from magnetic excitations in $Bi_2Sr_2CaCu_2O_{8+\delta}$. *Nature* **398,** 588 (1999).
13. Capogna, L. et al. Odd and even magnetic resonant modes in highly overdoped $Bi_2Sr_2CaCu_2O_{8+\delta}$. *Phys. Rev. B* **75,** 060502R (2007).
14. He, H. et al. Magnetic resonant mode in the single-layer high-temperature superconductor $Tl_2Ba_2CuO_{6+\delta}$. *Science* **295,** 1045 (2002).
15. Yu, G. et al. Magnetic resonance in the model high-temperature superconductor $HgBa_2CuO_{4+\delta}$. arXiv:0810.5759v1.
16. Li, Y. et al. manuscript in preparation.
17. Wilson, S. D. et al. Resonance in the electron-doped high-transition-temperature superconductor $Pr_{0.88}LaCe_{0.12}CuO_{4-\delta}$. *Nature* **442,** 59 (2006).
18. Yu, G. et al. Two characteristic energies in the low-energy antiferromagnetic response of the electron-doped high-temperature superconductor $Nd_{2-x}Ce_xCuO_{4+\delta}$. arXiv:0803.3250v2.
19. Christensen, N. B. et al. Dispersive excitations in the high-temperature superconductor $La_{2-x}Sr_xCuO_4$. *Phys. Rev. Lett.* **93,** 147002 (2004).
20. Lipscombe, O. P. et al. Persistence of high-frequency spin fluctuations in overdoped superconducting $La_{2-x}Sr_xCuO_4$ ($x$=0.22). *Phys. Rev. Lett.* **99,** 067002 (2007).
21. Fong, H. F. et al. Spin susceptibility in underdoped $YBa_2Cu_3O_{6+x}$. *Phys. Rev. B* **61,** 14773 (2000).
22. Hüfner, S., Hossain, M. A., Damascelli, A. & Sawatzky, G. A. Two gaps make a high-temperature superconductor? *Rep. Prog. Phys.* **71,** 062501 (2008).

23. Lee, W. *et al*. Abrupt onset of a second energy gap at the superconducting transition of underdoped Bi2212. *Nature* **450,** 81 (2007).
24. Fischer, Ø., Kugler, M., Maggio-Aprile, I. & Berthod, C. Scanning tunneling spectroscopy of high-temperature superconductors. *Rev. Mod. Phys.* **79,** 353 (2007).
25. Norman, M. R., Pines, D. & Kallin, C. The pseudogap: friend or foe of high $T_c$? *Adv. Phys.* **54,** 715 (2005).
26. Hao, Z. & Chubukov, A.V. Magnetic resonance in the cuprates - exciton, plasmon, or π-mode. arXiv:0812.2697v1.
27. Brehm, S., Arrigoni, E., Aichhorn, M. & Hanke, W. Theory of two-particle excitations and the magnetic susceptibility in high-Tc cuprate superconductors. arXiv:0811.0552v3.
28. Yang, J. *et al*. Exchange boson dynamics in cuprates: optical conductivity of $HgBa_2CuO_{4+\delta}$. *Phys. Rev. Lett.* **102,** 027003 (2009).
29. van Heumen, E. *et al*. Observation of a robust peak in the glue function of the high-Tc cuprates in the 50-60 meV range. arXiv:0807.1730v3.
30. Kee, H.-Y., Kivelson, S. A. & Aeppli, G. Spin-1 neutron resonance peak cannot account for electronic anomalies in the cuprate superconductors. *Phys. Rev. Lett.* **88,** 257002 (2002).
31. Kyung, B., Sénéchal, D. & Tremblay, A.-M. S. Retarded interactions, short-range spin fluctuations, and high temperature superconductivity. arXiv:0812.1228v1.

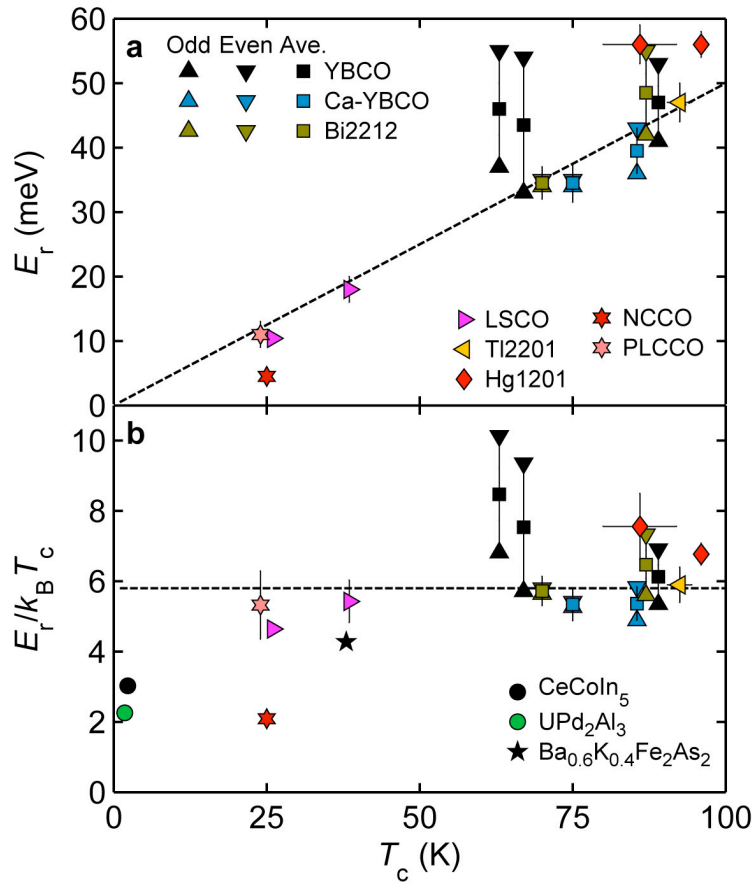

**Figure 1 | Resonance energy and ratio $E_r/T_c$ versus $T_c$. a,** Resonance energy versus $T_c$ for cuprates. For the double-layer hole-doped cuprates (YBCO, Ca-doped YBCO, and Bi2212), up (down) triangles indicate odd (even) resonance modes, while squares represent the average of the two mode energies[9-11,13,21]. For LSCO, the resonance is absent and the characteristic energy shown is the maximum in the local magnetic susceptibility, defined as the integral of $\chi''(\mathbf{Q}, \omega)$ over the Brillouin zone[19-20]. The interpretation of neutron data for the magnetic resonance in the electron-doped cuprates PLCCO[17] and NCCO[18] is discussed in the Supplementary Information. The dashed line indicates $E_r = 5.8\ k_B T_c$. The results for Hg1201[15-16], NCCO[18], and the most underdoped YBCO samples[11,21] deviate strongly from this relationship. Figure S1 shows the estimated doping dependence of $E_r$ for the hole-doped cuprates, including additional results for the odd-parity mode. **b,** The ratio $E_r/T_c$ versus $T_c$ for all three types of unconventional superconductors. In addition to the results in (**a**), this figure includes those for the heavy-fermion superconductors UPd$_2$Al$_3$[3] and CeCoIn$_5$[4] as well as for the iron arsenide (Ba,K)Fe$_2$As$_2$[5]. All three exhibit a rather small value of $E_r/T_c$. The resonance energy and $T_c$ values are summarized in Table S1 in the Supplementary Information.

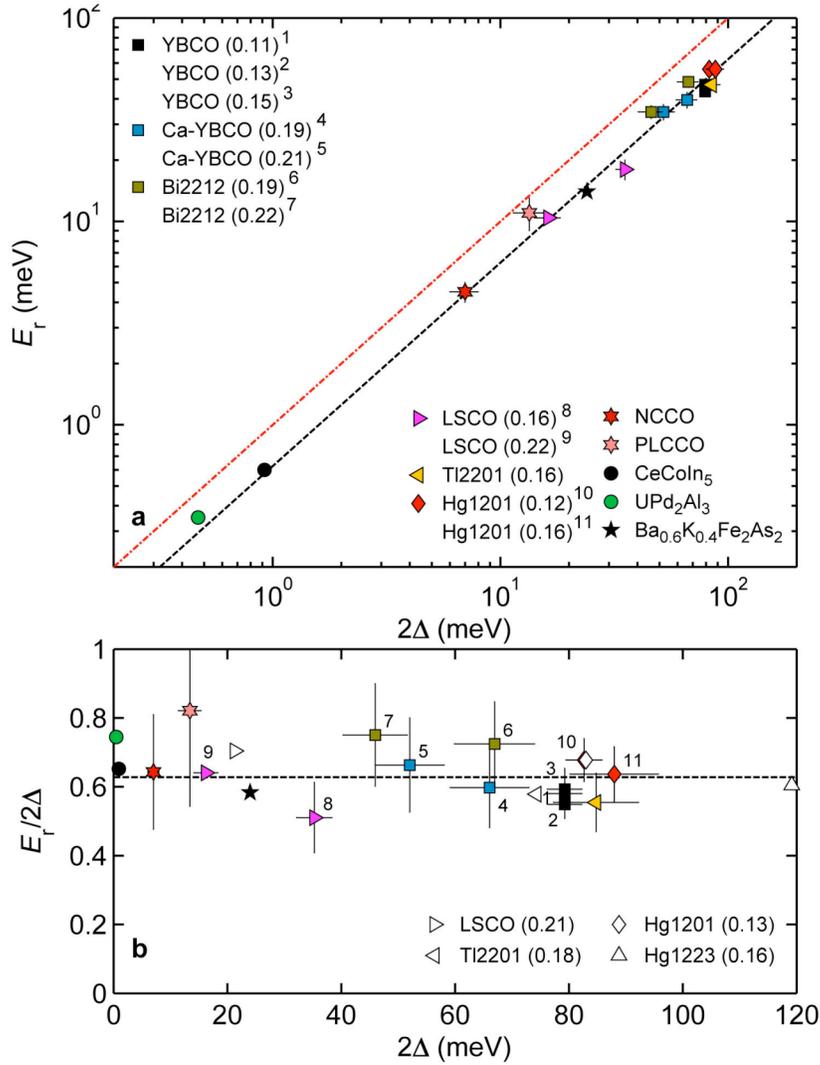

**Figure 2 | Resonance energy and ratio $E_r/2\Delta$ versus $2\Delta$. a,** The resonance energy $E_r$ versus the pair-breaking energy $2\Delta$ (twice the SC gap maximum). $E_r$ of the double-layer cuprates is represented by the average of the even and odd parity mode energies. The black dashed line is a fit to all of the data: $E_r/2\Delta = 0.64(2)$. Unlike the case in Fig. 1, the results for Hg1201[15-16], NCCO[18], and the average mode energy of the most underdoped YBCO samples[11,21] all follow this relationship. As a reference, the red dot-dashed line represents $E_r = 2\Delta$. Figure S2 shows the resonance energy as a function of $2\Delta$ on a linear scale. **b,** The ratio of $E_r/2\Delta$ versus $2\Delta$. The dashed horizontal line indicates the value $E_r/2\Delta = 0.64$. The open symbols indicate the maximum in the bosonic spectral function ($E_{opt}$) extracted from optical spectroscopy for the single-layer cuprates LSCO, Tl2201, and Hg1201, and for triple-layer Hg1223[28]. For the hole-doped cuprates, estimated hole concentrations are shown in parentheses. The $2\Delta$ values and hole concentrations are obtained as discussed in the Supplementary Information. The characteristic energy and SC gap values are summarized in Table S1 (neutron scattering results) and Table S2 (optical conductivity results).

**Supplementary Information for "Universal relation between magnetic resonance and superconducting gap in unconventional superconductors"**

G. Yu, Y. Li, E. M. Motoyama and M. Greven

**Determination of superconducting gap of hole-doped cuprates**.
Figure S3 describes how we estimate $\Delta$ for the hole-doped cuprates. For all other compounds, the gap values were taken directly from the literature, as summarized in Table S1. For the hole-doped cuprates, there is increasing evidence for the existence of two gaps[22]. The first gap, $\Delta_0$, is defined in the nodal region with the $d$-wave form $\Delta_d(k) = \Delta_0 |\cos(k_x)-\cos(k_y)|/2$; it is generally considered to be the SC gap, and we use $\Delta = \Delta_0$. The second gap, $\Delta_1$, is found in the antinodal region[S1] and is referred to the "pseudogap"; it increases with decreasing doping and has been associated with many unusual normal state properties observed below a characteristic temperature $T^{*}$[25]. In the overdoped regime, photoemission measurements reveal that the momentum dependence of the gap function is well described by the $d$-wave form along the whole Fermi surface, indicating that the two gaps are identical[23] or that the pseudogap is smaller than the SC gap and therefore invisible. This is supported by the good agreement between the gap values obtained from thermal conductivity measurements, which probe the nodal region, and from superconductor-insulator-superconductor (SIS) tunneling and recent STM measurements[S2-S11], which probe the antinodal region. As shown in Fig. S3a, we find for the overdoped region that the SC gap of single- and double-layer cuprates can be approximated by a universal linear function of doping: $2\Delta_0/k_B T_c = a(p-p_0)$, with $a = 53(7)$ and $p_0 = 0.36(1)$. In the underdoped regime, the momentum dependence of the gap function deviates from the $d$-wave form, and Fig. S3b shows that the two gaps differ[23,S12-S13]. Very recent photoemission experiments reveal that $2\Delta_0$ does not scale with $T_c$, but instead is nearly independent of doping for $0.08 < p < 0.16$[23,S12-S13]. Here we approximate the nodal gap with a constant $2\Delta_0/k_B T_c^{op} = 9.9(4)$, where $T_c^{op}$ is the transition temperature at optimal doping ($p = 0.16$). This estimate is used for underdoped Hg1201 and YBCO. The values of $2\Delta_0$ obtained in the two doping regimes agree at optimal doping. We exclude early photoemission and STM work in our estimate due to the large scatter in the experimental gap values.

The presence of two gaps is also seen from Raman scattering, where the responses in $B_{2g}$ and $B_{1g}$ symmetry mainly stem from the nodal and antinodal regions, respectively. However, quantitative extraction of the gap values is still difficult[S14]. A recent analysis of the $B_{2g}$ Raman responses in YBCO and Bi2212 suggests that the SC gap function can be described by a $d$-wave form, and that the gap is proportional to 9 $k_B T_c$ over a wide doping range[S15]. While this estimate agrees rather well with the gap values from other techniques at and above optimal doping (Fig. S3b), it is inconsistent with the nearly doping independence of the nodal gap suggested by the recent photoemission work for Bi2212 and LSCO[23,S12-S13]. The origin of this discrepancy is still unclear. However, we note that Raman scattering is sensitive to the coherent quasiparticles[S14], and that at low doping, the coherent quasi-particle spectral weight diminishes quickly away from the nodes. This may lead to an underestimate of the genuine gap value, as the Raman response might be restricted to a very narrow region around the nodes.

We emphasize that $\Delta_0$ is a robust characteristic energy on the underdoped side since it corresponds to the energy scale of a distinct 'kink' in the momentum dependence of the gap function seen in photoemission[23,S12-S13]. In addition, for Bi2212, STM reveals a 'kink' feature in the spectra below the gap energy $\Delta_1$ which may naturally be explained by the kink in the gap function. While the STM kink energy is smaller than the value of $\Delta_0$ shown in Fig. S2b, it scales linearly with $\Delta_0$ over a wide range of doping[S11].

We note that it has been observed that the magnetic resonance in Bi2212 shows significant broadening in energy[S16], in contrast to the energy-resolution limited peak in YBCO[10], Tl2201[14], and Hg1201[15]. This broadening can be attributed to electronic inhomogeneities or defects in Bi2212. A comparison of the energy width of the resonance with the width of the spatial gap distribution reveals an approximately linear correlation between the two quantities[S16]. This observation was interpreted in the context of excitonic models, in which the resonance energy is affected by the gap size[2]. The proportionality observed in Ref. S16 is more naturally derived from the universal scaling relation $E_r \propto 2\Delta$ found here.

**Magnetic resonance in the electron-doped cuprates**.
The first report suggesting the existence of a magnetic resonance in the electron-doped curpates was Ref. 14 for PLCCO. Upon cooling into the SC state, an enhancement of the staggered susceptibility was found near 11 meV. This value is consistent with the linear relation $E_r = 5.8\,k_B T_c$, which was taken as evidence for universality across the hole- and electron-doped cuprates. A measurement of NCCO revealed a similar enhancement around 9-10 meV[S17]. However, it was subsequently found for NCCO that the enhancement at these energies is very broad and attributable to a general spectral weight shift from below $2\Delta$ to higher energies[18]. Moreover, a rather well-defined susceptibility enhancement was found below $2\Delta$ (at about 4.5 meV). We use this newer and lower characteristic energy scale for NCCO in our analysis. For NCCO, the SC gap has been obtained from Raman scattering[S18], STM[S19], and photoemission[S20] experiments, and the three methods give consistent results, with $2\Delta$ close to but less than 10 meV near optimal doping. For PLCCO, on the other hand, recent STM work suggests a rather large value of $2\Delta \sim 13$ meV near optimal doping ($x = 0.12$; $T_c = 24$ K)[S21], whereas photoemission results indicate a value much less than 10 meV ($x = 0.11$; $T_c = 26$ K)[S22]. Since the STM work was done on samples of the same composition and origin as the crystal studied by neutron scattering[17], we use the former (larger) value. As in the case of NCCO, it is possible that the 11 meV feature is not a magnetic resonance.

**Iron arsenide superconductors**.
New results for the iron arsenide superconductors continue to be announced. A recent neutron measurement reveals a resonance at $E_r = 9.6$ meV in electron-doped $BaFe_{1.84}Co_{0.16}As_2$ ($T_c = 22$ K)[S23], which has the same crystal structure as hole-doped $(Ba,K)Fe_2As_2$. A resonance has also been observed in $BaFe_{1.9}Ni_{0.1}As_2$ ($T_c = 20$ K)[S24-S25]. Interestingly, the resonance in this compound is found to disperse along the c-axis, with values of 9 meV and 7 meV at $\mathbf{Q} = (\pi/a, \pi/a, 0)$ and $Q = (\pi/a, \pi/a, -\pi/c)$, respectively[S24]. Recent ARPES results on BaFe$_{1.84}$Co$_{0.16}$As$_2$ suggest a three-dimensional Fermi surface[S26], which is further supported by measurements of the upper critical field[S27-S28]. It follows naturally that the SC gap in these compounds may also depend on the momentum transfer along the $c$-axis and that, in principle, the resonance energy will scale with the superconduting gap. We are not able to confirm the universal scaling in these samples because the gap has not yet been reported for these Co and Ni concentrations. Regardless of the dimensionality of the Fermi surface and the symmetry of the order parameter, the resonance in the iron arsenide compounds seems to appear for the same reason as in $d$-wave superconductors if, indeed, the gaps along the hole and electron Fermi surfaces have different signs.

**References**

S1. Damascelli, A., Hussain, Z. & Shen, Z.-X. Angle-resolved photoemission studies of the cuprate superconductors. *Rev. Mod. Phys.* **75,** 473 (2003).
S2. Sutherland, M. *et al*. Thermal conductivity across the phase diagram of cuprates: Low-energy quasiparticles and doping dependence of the superconducting gap. *Phys. Rev. B* **67,** 174520 (2003).
S3. Hawthorn, D. G. *et al*. Doping dependence of the superconducting gap in $Tl_2Ba_2CuO_{6+\delta}$ from heat transport. *Phys. Rev. B* **75,** 104518 (2007).
S4. Ozyuzer, L. *et al*. Tunneling spectroscopy of $Tl_2Ba_2CuO_6$. *Physica C* **320,** 9 (1999).
S5. Momono, N. *et al*. Energy gap evolution over wide temperature and doping ranges in $Bi_2Sr_2CaCu_2O_{8+\delta}$. *Mod. Phys. Lett. B* **17,** 401 (2003).
S6. Miyakawa, N. *et al*. Strong Dependence of the superconducting gap on oxygen doping from tunneling measurements on $Bi_2Sr_2CaCu_2O_{8-\delta}$. *Phys. Rev. Lett.* **80,** 157 (1998).
S7. Miyakawa, N. *et al*. Predominantly superconducting origin of large energy gaps in underdoped $Bi_2Sr_2CaCu_2O_{8+\delta}$ from tunneling spectroscopy. *Phys. Rev. Lett.* **83,** 1018 (1999).
S8. McElroy, K. *et al*. Coincidence of checkerboard charge order and antinodal state decoherence in strongly underdoped superconducting $Bi_2Sr_2CaCu_2O_{8+\delta}$. *Phys. Rev. Lett.* **94,** 197005 (2005).
S9. Lee, J. *et al*. Interplay of electron-lattice interactions and superconductivity in $Bi_2Sr_2CaCu_2O_{8+\delta}$. *Nature* **442,** 546 (2006).
S10. Gomes, K. K. *et al*. Visualizing pair formation on the atomic scale in the high-$T_c$ superconductor $Bi_2Sr_2CaCu_2O_{8+\delta}$. *Nature* **447,** 569 (2007).
S11. Alldredge, J. W. *et al*. Evolution of the electronic excitation spectrum with strongly diminishing hole density in superconducting $Bi_2Sr_2CaCu_2O_{8+\delta}$. *Nature Physics* **4,** 319 (2008).
S12. Yoshida, T. *et al*. Universal versus material-dependent two-gap behaviors in the high-$T_c$ cuprates: Angle-resolved photoemission study of $La_{2-x}Sr_xCuO_4$. arXiv:0812.0155v1.
S13. Tanaka, K. *et al*. Distinct Fermi-momentum-dependent energy gaps in deeply underdoped Bi2212. *Science* **314,** 1910 (2006).
S14. Devereaux, T. P. & Hackl, R. Inelastic light scattering from correlated electrons. *Rev. Mod. Phys.* **79,** 175 (2007).
S15. Munnikes, N. *et al*. Pair breaking versus symmetry breaking: Origin of the Raman modes in superconducting cuprates. arXiv:0901.3448v1.
S16. Fauqué, B. *et al*. Dispersion of the odd magnetic resonant mode in near-optimally doped $Bi_2Sr_2CaCu_2O_{8+\delta}$. *Phys. Rev. B* **76,** 214512 (2007).
S17. Zhao, J. *et al*. Neutron-spin resonance in the optimally electron-doped superconductor $Nd_{1.85}Ce_{0.15}CuO_{4-\delta}$. *Phys. Rev. Lett.* **99,** 017001 (2007).
S18. Qazilbash, M. M. *et al*. Evolution of superconductivity in electron-doped cuprates: Magneto-Raman spectroscopy. *Phys. Rev. B* **72,** 214510 (2005).

S19. Kashiwaya, S. *et al*. Tunneling spectroscopy of superconducting $Nd_{1.85}Ce_{0.15}CuO_{4-\delta}$. *Phys. Rev. B* **57,** 8680 (1998).
S20. Sato, T., Kamiyama, T., Takahashi, T., Kurahashi, K. & Yamada, K. Observation of $d_{x^2-y^2}$-like superconducting gap in an electron-doped high-temperature superconductor. *Science* **291,** 1517 (2001).
S21. Niestemski, F. C. *et al*. A distinct bosonic mode in an electron-doped high-transition-temperature superconductor. *Nature* **450,** 1058 (2007).
S22. Matsui, H. *et al*. Direct observation of a nonmonotonic $d_{x^2-y^2}$-wave superconducting gap in the electron-doped high-$T_c$ superconductor $Pr_{0.89}LaCe_{0.11}CuO_4$. *Phys. Rev. Lett.* **95,** 017003 (2005).
S23. Lumsden, M. D. *et al*. Two-dimensional resonant magnetic excitation in $BaFe_{1.84}Co_{0.16}As_2$. arXiv:0811.4755v2.
S24. Chi, S. *et al*. Three-dimensional Resonance in superconducting $BaFe_{1.9}Ni_{0.1}As_2$. arXiv:0812.1354v1.
S25. Li, S. *et al*. Spin gap and magnetic resonance in superconducting $BaFe_{1.9}Ni_{0.1}As_2$. arXiv:0902.0813v1.
S26. Vilmercati, P. *et al*. Evidence for three-dimensionality in the Fermi surface topology of layered electron doped $Ba(Fe_{1-x}Co_x)_2As_2$ iron superconductors. arXiv:0902.0756.
S27. Altarawneh, W. W. *et al*. Determination of anisotropic $H_{c2}$ up to 60 T in $Ba_{0.55}K_{0.45}Fe_2As_2$ single crystals. *Phys. Rev. B* **78**, 220505 (2008).
S28. Yuan, H. Q. *et al*. Nearly isotropic superconductivity in $(Ba,K)Fe_2As_2$. *Nature* **457**, 565 (2009).
S29. Liang, R., Bonn, D. A. & Hardy, W. N. Evaluation of $CuO_2$ plane hole doping in $YBa_2Cu_3O_{6+x}$ single crystals. *Phys. Rev. B* **73,** 180505(R) (2006).
S30. Tallon, J. L. *et al*. Generic superconducting phase behavior in high-$T_c$ cuprates: $T_c$ variation with hole concentration in $YBa_2Cu_3O_{7-\delta}$. *Phys. Rev. B* **51,** 12911 (1995).
S31. Mook, H. A. *et al*. Polarized neutron determination of the magnetic excitations in $YBa_2Cu_3O_7$. *Phys. Rev. Lett.* **70,** 3490 (1993).
S32. Fong, H. F. *et al*. Phonon and magnetic neutron scattering at 41 meV in $YBa_2Cu_3O_7$. *Phys. Rev. Lett.* **75,** 316 (1995).
S33. Bourges, P., Regnault, L. P., Sidis, Y. & Vettier, C. Inelastic-neutron-scattering study of antiferromagnetic fluctuations in $YBa_2Cu_3O_{6.97}$. *Phys. Rev. B* **53,** 876 (1996).
S34. Dai, P., Mook, H. A., Hunt, R. D. & Dogan, F. Evolution of the resonance and incommensurate spin fluctuations in superconducting $YBa_2Cu_3O_{6+x}$. *Phys. Rev. B* **63,** 054525 (2001).
S35. Stock, C. *et al*. Dynamic stripes and resonance in the superconducting and normal phases of $YBa_2Cu_3O_{6.5}$ ortho-II superconductor. *Phys. Rev. B* **69,** 014502 (2004).
S36. He, H. *et al*. Resonant spin excitation in an overdoped high temperature superconductor. *Phys. Rev. Lett.* **86,** 1610 (2001).
S37. Jourdan, M., Huth, M. & Adrin, H. Superconductivity mediated by spin fluctuations in the heavy-fermion compound $UPd_2Al_3$. *Nature* **398,** 47 (1999).
S38. Ding, H. *et al*. Observation of Fermi-surface–dependent nodeless superconducting gaps in $Ba_{0.6}K_{0.4}Fe_2As_2$. *Europhys. Lett.* **83,** 47001 (2008).
S39. Hwang, J. *et al*. Bosonic spectral density of epitaxial thin-film $La_{1.83}Sr_{0.17}CuO_4$ superconductors from infrared conductivity measurements. *Phys. Rev. Lett.* **100,** 137005 (2008).

S40. Schachinger, E. & Carbotte, J. P. Coupling to spin fluctuations from conductivity scattering rates. *Phys. Rev. B* **62,** 9054 (2000).

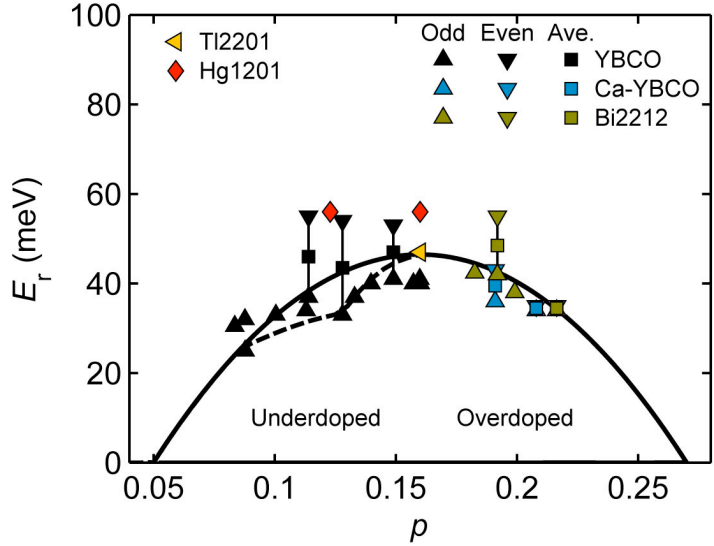

**Figure S1 | Doping dependence of the resonance energy for the hole-doped cuprates.** The hole concentration for YBCO is estimated from the doping dependence of $T_c$ established in Ref. S29 for YBCO, and the dashed line indicates $5.8\,k_B T_c$. For all the other cuprates, the hole concentration is estimated from the empirical formula $T_c/T_c^{op} = 1-82.6(p-0.16)^2$ (Ref. S30), where $T_c^{op}$ is the transition temperature at optimal doping ($p = 0.16$); the continuous line indicates $5.8\,k_B T_c$. The two estimates agree at low hole concentrations and on the overdoped side of the phase diagram. For the double-layer compounds YBCO[10-11,21], Ca-YBCO[9,11], and Bi2212[13], the up and down triangles represent the odd and even resonance modes, respectively, while the squares indicate the average of the two. Additional results for only the odd-parity mode of YBCO[S31-S35] and Bi2212[12,S36] are shown.

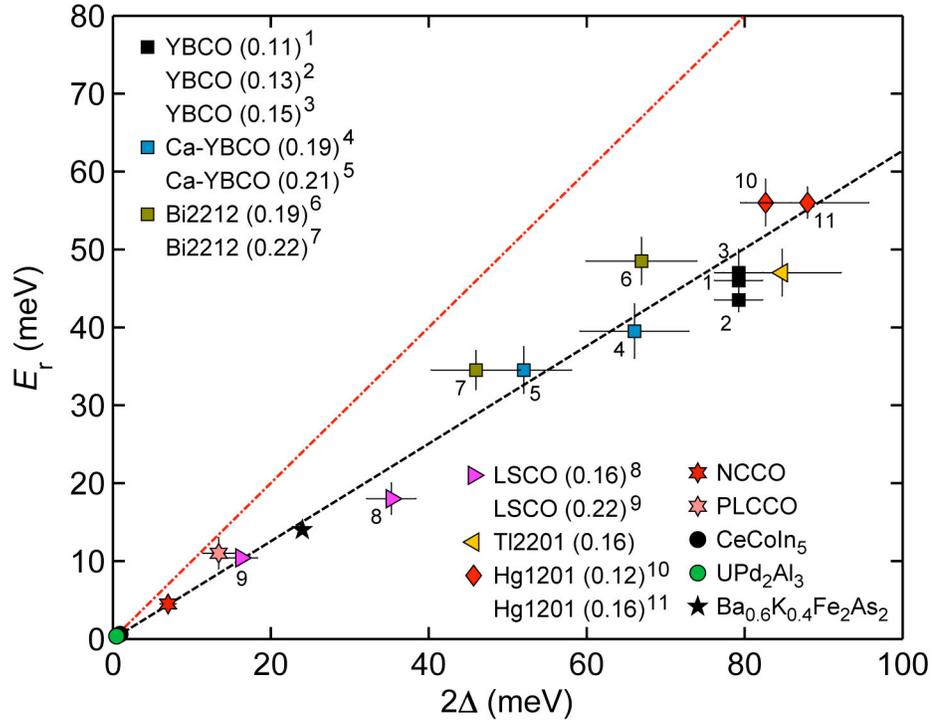

**Figure S2 | The magnetic resonance energy obtained from neutron scattering as a function of the SC gap 2Δ on a linear scale.** The black line indicates $E_r/2\Delta = 0.64$, whereas the red reference line indicates $E_r/2\Delta = 1$.

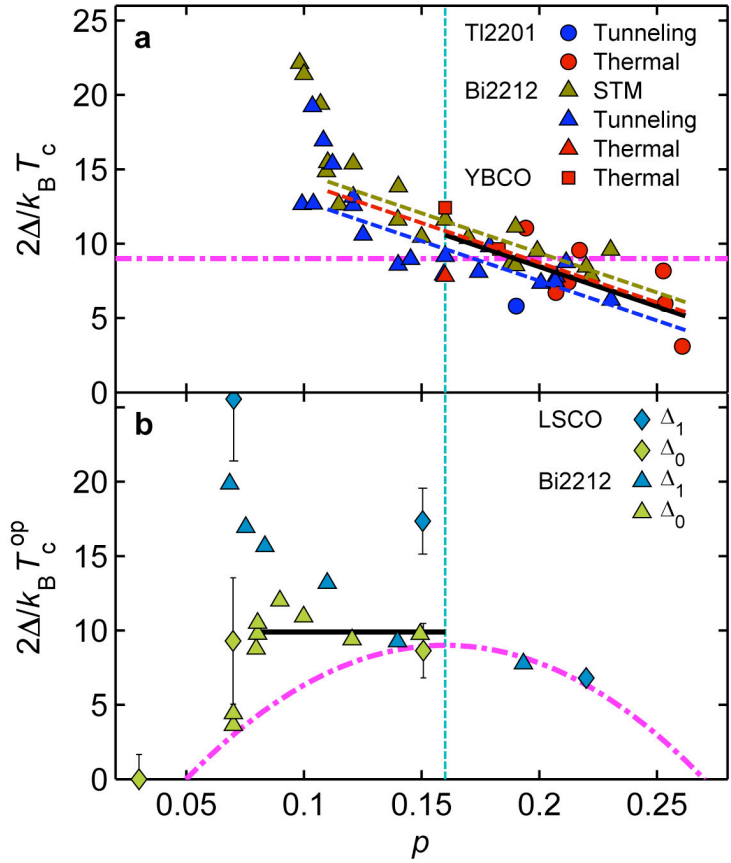

**Figure S3 | Estimation of the pair-breaking energy 2Δ for the hole-doped cuprates. a,** Nodal gap ($\Delta_0$) from thermal conductivity measurements[S2-S3], and antinodal gap ($\Delta_1$) from tunneling measurements[S4-S7] and STM[S8-S11] for several hole-doped cuprates. The dashed lines are linear fits of the data with the same slope but different offset for different experimental probes. The solid black line is the average of the three linear fits. **b,** $\Delta_0$ and $\Delta_1$ from ARPES measurements for LSCO[S12] and Bi2212[23,S13]. Above optimal doping ($p >$ 0.16), $\Delta_0 \approx \Delta_1$. For hole concentrations $p$ between 0.08 and 0.15, $\Delta_0$ is approximately constant. The solid horizontal line is a fit of $\Delta_0$ to a constant in the doping range $p$ = 0.08-0.15. The estimated values of $\Delta_0$ (black lines) in (**a**) and (**b**) are equal at $p$ = 0.16 (vertical dashed line). The purple dot-dashed line in (**a**) and (**b**) indicates the relation $2\Delta = 9\, k_B T_c$ as described in the text. The estimated gap values in (**b**) are used for the two most underdoped YBCO samples[11,21] ($p \approx$ 0.11, 0.13 and 0.15), for underdoped Hg1201[16] ($p \approx$ 0.12), and for the underdoped Hg1201 ($p \approx$ 0.13) sample studied by optical conductivity[28]. All other gap estimates for the hole-doped cuprates are based on (**a**). The hole concentrations are estimated as described in Fig. S1.

| Compound | Doping | $T_c$ (K) | $E_r$ (meV) | Ref. | $2\Delta$ (meV) | Ref. | $E_r/2\Delta$ |
|---|---|---|---|---|---|---|---|
| $YBa_2Cu_3O_{6.6}$ | UD | 63 | 37(2), 55(6) | 11 | 79(3) | | 0.58(7) |
| $YBa_2Cu_3O_{6.7}$ | UD | 67 | 33, 54 | 21 | 79(3) | | 0.55 |
| $YBa_2Cu_3O_{6.85}$ | UD | 89 | 41(1), 53(4) | 10 | 79(3) | | 0.59(5) |
| $Y_{0.9}Ca_{0.1}Ba_2Cu_3O_7$ | OD | 85.5 | 36(2), 43(5) | 9 | 66(8) | | 0.60(12) |
| $Y_{0.85}Ca_{0.15}Ba_2Cu_3O_7$ | OD | 75 | 34(3), 35(3) | 11 | 52(7) | | 0.66(14) |
| $Bi_2Sr_2CaCu_2O_{8+\delta}$ | OD | 87 | 42(3), 55(3) | 13 | 66(8) | | 0.72(12) |
| $Bi_2Sr_2CaCu_2O_{8+\delta}$ | OD | 70 | 34(2), 35(3) | 13 | 46(6) | | 0.75(14) |
| $La_{2-x}Sr_xCuO_4$ | OP | 38.5 | 18(2) | 19 | 35(3) | | 0.51(10) |
| $La_{2-x}Sr_xCuO_4$ | OD | 26 | 10 | 20 | 16(2) | | 0.64 |
| $Tl_2Ba_2CuO_{6+\delta}$ | OP | 92.5 | 47(3) | 14 | 85(7) | | 0.55(9) |
| $HgBa_2CuO_{4+\delta}$ | UD | 86 | 56(3) | 16 | 83(3) | | 0.68(6) |
| $HgBa_2CuO_{4+\delta}$ | OP | 96 | 56(2) | 15 | 88(8) | | 0.64(8) |
| $Pr_{0.88}LaCe_{0.12}CuO_{4+\delta}$ | | 24 | 11(2) | 17 | 13(2) | S21 | 0.82 (28) |
| $Nd_{1.845}Ce_{0.155}CuO_{4+\delta}$ | | 25 | 4.5(5) | 18 | 7(1) | S18-S20 | 0.64(17) |
| $UPd_2Al_3$ | | 1.8 | 0.35 | 3 | 0.47 | S37 | 0.74 |
| $CeCoIn_5$ | | 2.3 | 0.60(3) | 4 | 0.92 | 4 | 0.65 |
| $Ba_{0.6}K_{0.4}Fe_2As_2$ | | 38 | 14 | 5 | 24 | S38 | 0.58 |

**Table S1 | Summary of SC transition temperature, magnetic resonance energy from neutron scattering, SC gap, and ratio $E_r/2\Delta$.** For the double-layer hole-doped compounds, we only list the results from experiments in which both even and odd parity modes were obtained. For the hole-doped cuprates, the values refer to the nodal gap $2\Delta_0$ and were obtained as described in the discussion of Fig. S3. For hole-doped cuprates: UD (underdoped), OP (optimally doped), OD (overdoped). The uncertainties in $E_r$ and $2\Delta$ are provided as listed in the respective reference or as obtained from Fig. S3. In those cases in which no error estimate is available for $E_r$, we do not attempt to provide an error estimate for the ratio ratio $E_r/2\Delta$.

| Compound | Doping | $T_c$ (K) | $E_{opt}$ (meV) | Ref. | $2\Delta$ (meV) | $E_{opt}/2\Delta$ |
|---|---|---|---|---|---|---|
| $La_{2-x}Sr_xCuO_4$ | OD | 31 | 15 | 28,S39 | 21 | 0.71 |
| $Tl_2Ba_2CuO_{6+\delta}$ | OD | 90 | 43 | 28,S40 | 73 | 0.59 |
| $HgBa_2CuO_{4+\delta}$ | UD | 90 | 56 | 28 | 84 | 0.70 |
| $HgBa_2Ca_2Cu_3O_{8+\delta}$ | OD | 130 | 72 | 28 | 117 | 0.62 |

**Table S2 | Summary of SC transition temperature, characteristic energy from optical spectroscopy, SC gap, and ratio $E_{opt}/2\Delta$ for four single and triple-layer hole-doped cuprates[28,S39-S40].** The SC gap was estimated as described in the discussion of Fig. S3. Since estimates of the error in $E_{opt}$ are unavailable, we do not provide error estimates for $2\Delta$ and $E_{opt}/2\Delta$.